\documentclass{article}
\usepackage[final]{neur}
\usepackage[utf8]{inputenc} 
\usepackage[T1]{fontenc}    
\usepackage{hyperref}       
\usepackage{url}            
\usepackage{booktabs}       
\usepackage{amsfonts}       
\usepackage{nicefrac}       
\usepackage{microtype}      
\usepackage{lipsum}
\usepackage{amsmath}
\usepackage{stackengine}
\usepackage{float}
\usepackage{amssymb}
\usepackage{graphicx}
\graphicspath{{Images/}}
\usepackage{subfig}
\usepackage{float}
\usepackage{geometry}
\usepackage{color, colortbl}
\usepackage{rotating}
\usepackage{subfig}
\usepackage{tikz}
\usepackage{xcolor}
\usepackage{animate}
\newcommand{\vect}[1]{\boldsymbol{#1}}
\hypersetup{
    colorlinks,
    linkcolor={red!50!black},
    citecolor={blue!80!black},
    urlcolor={blue!80!black}
}
\usepackage[sorting=none,style=ieee]{biblatex} 
\bibliography{references} 

\title{Functional Correspondences in the Human and Marmoset Visual Cortex During Movie Watching: Insights from Correlation, Redundancy, and Synergy}

\author{%
  Qiang Li$^{1}$, Ting Xu$^{2}$, Vince D. Calhoun$^{1}$\\
  $^{1}$Tri-Institutional Center for Translational Research in Neuroimaging and Data Science (TReNDS)\\
  Georgia State University, Georgia Institute of Technology, Emory University, Atlanta, GA, United States\\
  $^{2}$Child Mind Institute, Center for the Integrative Developmental Neuroscience, New York, NY, United States}
  
\begin{document}
\maketitle
\begin{abstract}
The world of beauty is deeply connected to the visual cortex, as perception often begins with vision in both humans and marmosets. Quantifying functional correspondences in the visual cortex across species can help us understand how information is processed in the primate visual cortex, while also providing deeper insights into human visual cortex functions through the study of marmosets. \textcolor{black}{In this study, to investigate their functional correspondences, we used 13 healthy human volunteers (9 males and 4 females, aged 22-56 years) and 8 common marmosets (6 males and 2 females, aged 20-42 months). We then measured pairwise and beyond-pairwise correlations, redundancy, and synergy in movie-driven fMRI data across species.} \textcolor{black}{First, we consistently observed a high degree of functional similarity in visual processing within and between species, suggesting that integrative processing mechanisms are preserved in both humans and marmosets, despite potential differences in their specific activity patterns.} \textcolor{black}{Second, we found that the strongest functional correspondences during movie watching occurred between the human peri-entorhinal and entorhinal cortex (PeEc) and the occipitotemporal high-level visual regions in the marmoset, reflecting a synergistic functional relationship. This suggests that these regions share complementary and integrated patterns of information processing across species.}  Third, redundancy measures maintained stable high-order hubs, indicating a steady core of shared information processing, while synergy measures revealed a dynamic shift from low- to high-level visual regions as interaction increased, reflecting adaptive integration. This highlights distinct patterns of information processing across the visual hierarchy. Ultimately, our results reveal the marmoset as a compelling model for investigating visual perception, distinguished by its remarkable functional parallels to the human visual cortex.
\end{abstract}


\section{Introduction}   
The visual system is one of the most powerful information-communication bridges between the complex redundancy world and the primate brain. Information processing in the primate visual brain is a key neuroscience and development problem~\cite{watanabe1960information,renyi1961measures,li2022functional,Varley2022MultivariateIT,li2022functionalnn}. \textcolor{black}{Since directly invasive recordings of human visual neural signals are not feasible, and the marmoset has become a widely used non-human primate model for studying human visual functions, it now serves as a key system for exploring the functional architecture of the human visual brain~\cite{Hori21PNAS,Qlinl24,Kell22iscience,Mitchell14jn,Ghahremani16,Dureux23}. Therefore, to understand how information processing in the human brain relates to that in the marmoset visual brain, we focused on regions showing cross-species functional correspondence. Previous research has shown that the visual cortex exhibits the greatest similarity across species~\cite{Hori21PNAS,Qlinl24,Kell22iscience}.}

The complex process of naturalistic information processing in the primate visual brain has been the subject of extensive research, yet many of the underlying neural mechanisms remain poorly understood. \textcolor{black}{Recent advances, however, have begun to clarify this picture. In particular, studies have identified structural and functional homologies between brain regions in humans and marmosets~\cite{Dureux23,Schaeffer20nc,Gilbert21nc}. These cross-species comparisons provide a valuable framework for exploring conserved visual processing strategies and for advancing our understanding of how the primate brain, including that of the marmoset, processes complex, naturalistic stimuli.}

Furthermore, to investigate functional correspondences between humans and marmosets during naturalistic movie viewing, particularly with complex and dynamic content such as faces, bodies, and scenes, we analyzed neural responses to identify shared visual processing mechanisms. This approach provides valuable insights into how these stimuli are represented in the brain across species and helps uncover both commonalities and differences in their underlying neural architectures. While previous studies have primarily focused on pairwise interactions to explain cross-species correspondence~\cite{Hori21PNAS}, these methods are limited because they capture only pairwise relationships and may miss higher-order dependencies~\cite{li2022functional,Varley2022MultivariateIT,li2022functionalnn}. Importantly, understanding synergistic information processing in the brain is essential, as this aspect is often overlooked when examining functional correspondence across species~\cite{luppi2022synergistic}.

To address these limitations, we employ both pairwise and higher-order correlation metrics, along with redundancy and synergy measures, to quantify functional similarities and differences in functional magnetic resonance imaging (fMRI) activity within the visual areas of humans and marmosets during movie watching. \textcolor{black}{This comprehensive approach offers new insights into the functional correspondences within the visual cortex of both species.} Our goal is to assess how visual information is processed by examining correlations, redundancy, and synergy among visual regions in response to naturalistic stimuli across species. \textcolor{black}{In particular, redundancy and synergy measures highlight distinct patterns of information integration within the primate visual cortex, offering complementary perspectives on cross-species functional organization.}

\section{Materials and Methods}
\subsection{Data Acquisition and Preprocessing}
We made use of existing open-source, movie-driven task fMRI datasets \cite{Hori21PNAS}, which include data from 13 healthy human volunteers (9 males and 4 females, aged 22-56 years) and 8 common marmosets (6 males and 2 females, aged 20-42 months). 

\textcolor{black}{Both humans and marmosets participated in a naturalistic movie-watching experiment to assess functional correspondences in the visual cortex. Marmosets underwent surgery to implant head chambers for stable head fixation during MRI scanning and were trained to acclimate to the MRI environment. Participants from both species watched a 15-minute silent movie, edited from the BBC documentary Hidden Kingdoms: Urban Jungles, with scenes selected to emphasize content relevant to marmosets. Human MRI data were acquired using a 7T Siemens scanner with a 32-channel receive coil, while marmoset data were collected using a 9.4T Bruker scanner with a custom-built gradient coil and a 5-channel receive coil. Eye tracking confirmed that marmosets remained awake during scanning. The movie was presented via PowerPoint on a MacBook Pro.} More details on the scanning parameters for both species can be found in~\cite{Hori21PNAS}. 

\begin{figure}[!ht]
    \centering
    \includegraphics[width=0.95\textwidth, height=12cm]{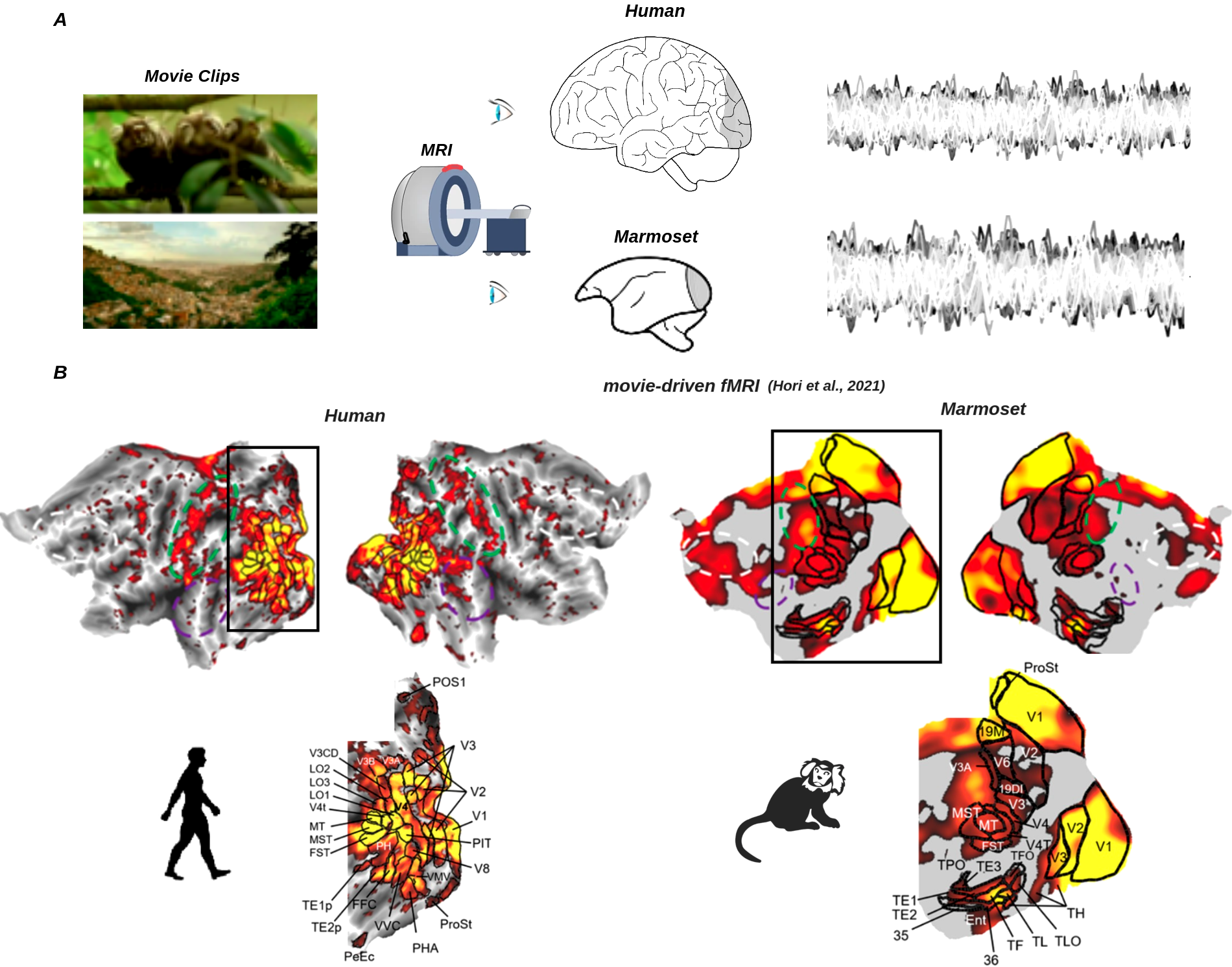} 
    \caption{The aligned activated visual cortex between humans and marmosets during naturalistic movie stimuli.
    \textit{\textbf{A.}} Naturalistic movie stimuli were presented to both humans and marmosets while brain activity was recorded using a MRI machine. After preprocessing, time courses from the activated visual regions were extracted.
    \textbf{\textit{B.}} The aligned activated visual cortex in humans and marmosets during naturalistic movie stimuli are presented. In the human visual cortex, 26 visual regions are labeled, while 25 visual regions are labeled in the marmoset visual cortex. The flat cortical maps of the human and marmoset brains were adapted from~\cite{Hori21PNAS}.}
    \label{fig:1}
\end{figure}

\textcolor{black}{After preprocessing (including motion and distortion correction, and spatial normalization), 26 visual areas in humans and 25 areas in marmosets were identified during naturalistic movie viewing, as illustrated in Fig.\ref{fig:1}.} These visual areas exhibited varying degrees of activation in response to the movie stimuli, particularly those involving faces, bodies, and scenes, across both the human and marmoset visual cortex. \textcolor{black}{Using the multimodal cortical parcellation atlas~\cite{Glasser2016} for humans and the Paxinos atlas~\cite{Paxinos2012} for marmosets, time courses were extracted from all activated regions of interest (ROIs) in both species for subsequent analysis.}

\textcolor{black}{To assess the quality of the fMRI time courses, we computed the temporal signal-to-noise ratio (tSNR) across all ROIs for both species. In humans, the average tSNR across 26 ROIs was 17.85, while in marmosets, the mean tSNR across 25 ROIs was 15.39. These values indicate acceptable signal quality for fMRI analysis in both species, supporting the reliability of the measured brain activity across regions.}

\subsection{Theoretical Analysis}
\subsubsection{Correlation}
\subsubsubsection{\textbf{Pairwise Relationship Measures with Cross-Correlation}}

Letting $x_{i}(t)$ and $y_{j}(t)$ represent the fMRI signals for two visual regions $x_{i}$ and $y_{j}$ at time point $t (t = 1, …, T)$, then cross-correlation~\cite{rabiner1978digital} will be denoted as,

\begin{equation}
{\mathbf{\rho}}_{cc}(x_{i}(t),y_{j}(t))={\rho }_{x_{i}(t)y_{j}(t)}(m)=\left\{\begin{array}{cc}{\sum }_{i=0}^{t-m-1}{x}_{i+m}(t){y}_{j}^{\ast }(t) & if\,m\,\ge \,0\\ {\rho }_{y_{j}(t)x_{i}(t)}(\,-\,m) & if\,m\, < \,0\end{array}\right.
\end{equation}

where ${y}_{j}^{\ast}$ represents the complex conjugate of ${y}_{j}$. Index $m$ represents the displacement between the two signals and is referred to as a lag or lead depending on whether it takes on a positive or negative value. The cross-correlation generates a vector of similarity measures for each value of $m$, whereas \textcolor{black}{instantaneous correlation} produces a single summary measure. This approach can be useful for identifying visual brain regions that are not functionally connected at the same moment but become functionally related after a temporal delay.

\subsubsubsection{\textbf{Multi-way Relationship Measures with Multi-Correlation}}
To quantify interactions beyond pairwise relationships in the data, we applied multi-correlation, which captures interactions among multiple variables, estimated through linear fitting of the pairwise data~\cite{Drezner1995MultirelationaCA}. 

\begin{equation}
1-\lambda(R')
\label{eq.multic}
\end{equation}

\noindent where $R'$ denotes a minor of $R$ and $\lambda$ represents the smallest eigenvalue of the visual connectivity matrix. The strength of linear relationships between the random variables is determined by this measure. To compute each multi-correlation measure for \(\vect{k}\)-way (\(\vect{k}>2\)) interactions among \(\vect{n}\) variables, we iterate over each set of indices used to extract minors of \(R\). For each minor, the multi-correlation is computed according to Eq.\ref{eq.multic}. In total, there are \(\binom{\vect{n}}{\vect{k}}\) possible sets of interactions from the input complex signals.

To represent high-order interactions in a pairwise manner, we first construct a hypergraph where nodes represent visual regions, and hyperedges capture multi-region interactions. Since hypergraphs encode complex relationships beyond pairwise connections, we convert the hypergraph into a clique, where each hyperedge is transformed into fully connected subgraphs. This conversion allows high-order interactions to be analyzed using traditional graph-based methods while preserving critical multi-region dependencies~\cite{Pickard23PLOS}.

\subsubsection{Redundancy and Synergy}
\subsubsubsection{\textbf{Partial Information Decomposition}}

The partial information decomposition (PID) reveals the mutual information that two source variables $X$ and $Y$ provide about a third target variable $Z$, $\mathbf{I}(X,Y; Z)$, can be decomposed into different types of information: information provided by one source but not the other (unique information), information provided by both sources separately (redundant information), or jointly by their combination (synergistic information)~\cite{Williams2010NonnegativeDO}. It can be mathematically expressed as,

\begin{equation}
\begin{aligned}
    \mathbf{I}(X, Y ; Z)=\operatorname{\mathbf{R}}(X, Y ; Z)+\operatorname{\mathbf{U}}(X ; Z \mid Y)+ \\
    \operatorname{\mathbf{U}}(Y ; Z \mid X)+\operatorname{\mathbf{S}}(X, Y ; Z)
\end{aligned}
\end{equation}

Where $\operatorname{\mathbf{R}}(X, Y; Z)$ refers to redundancy information, $\operatorname{\mathbf{U}}(X; Z \mid Y)$, $\operatorname{\mathbf{U}}(Y; Z \mid X)$ indicates unique information from Y and X, respectively, and $\operatorname{\mathbf{S}}(X, Y; Z)$ refers to synergy information. These different approaches offer a more detailed suite of metrics to quantify the complex system. However, limitations remain in applying this framework to dynamic system with temporal dependencies such as brain activation. 

\subsubsubsection{\textbf{Integrated Information Decomposition}}

To address this challenge and access the redundancy and synergy for brain data, the Integrated Information Decomposition ($\Phi$ID) was developed to extend PID to measure the time-delayed mutual information taking into account both past and currents states of brain signal~\cite{Barrett2011PracticalMO,Albers12csf,mediano2021extended}, i.e., $\mathbf{I}\left(X_{t-\tau}^i, X_{t-\tau}^j ; X_t^i, X_t^j\right)$, where $X_{t-\tau}^i, X_{t-\tau}^j$ refers to past states of brain signal and $X_t^i, X_t^j$ refers to current states. Redundancy, synergy, and unique information can be calculated under the $\Phi$ID framework. In this study, we utilized the Gaussian solver implemented in the Java information dynamics toolkit to compute all information-theoretic quantities using $\Phi$ID~\cite{Lizier14fr}.

\section{Results}
\subsection{\textcolor{black}{Intra-species Functional Correspondences in Humans and Marmosets}}
\begin{figure}[!ht]
    \centering
    \includegraphics[width=\textwidth, height=12cm]{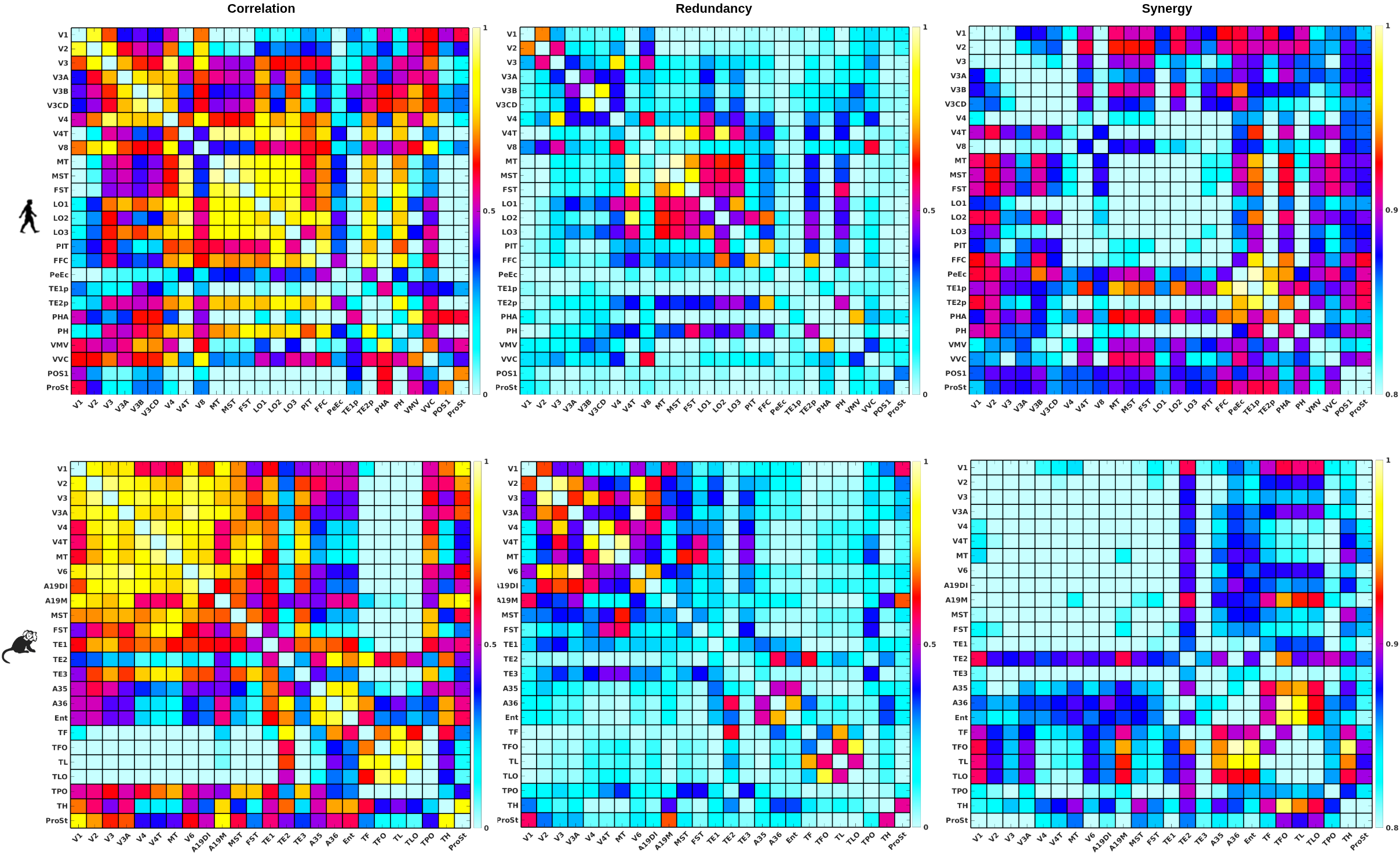} 
    \caption{\textcolor{black}{Intra-species functional correspondences were measured using correlation, redundancy, and synergy in both humans and marmosets. The top row presents the human visual cortex functional correspondence measures using correlation, redundancy, and synergy, respectively. The second row shows the corresponding functional correspondence measures for the marmoset visual cortex using the same metrics.}}
    \label{fig:1A}
\end{figure}

In Fig.\ref{fig:1A}, intra-species functional correspondences are presented. First, we observe that the correlation and redundancy measures reveal a high similarity in spatial connection patterns, whereas the synergy measure provides distinct connection information in both species. Specifically, in the human visual cortex, the perirhinal-entorhinal cortex (PeEc) and temporal area 1 posterior (TE1p) exhibit stronger functional synergy than other connections, which are not captured by the correlation and redundancy measures. In the marmoset visual cortex, temporal area TF, the occipital part (TFO), and area 36 (A36) show strong functional synergy connections relative to other regions. Finally, we observe that both species share similar connection patterns in the visual brain during movie-watching. For instance, the high-level visual cortex demonstrates strong functional synergy in both species, while the low- and middle-level visual regions show stronger connections than the high-level regions in redundancy measures. \textcolor{black}{These findings highlight a high degree of functional similarity in visual processing between humans and marmosets. Despite species-specific connection variations, both species exhibit comparable spatial connection structures and largely similar connection patterns during naturalistic visual stimulation, suggesting conserved mechanisms of visual information processing across primates.}

\begin{figure}[htbp]
    \centering
    \includegraphics[width=\textwidth, height=18cm]{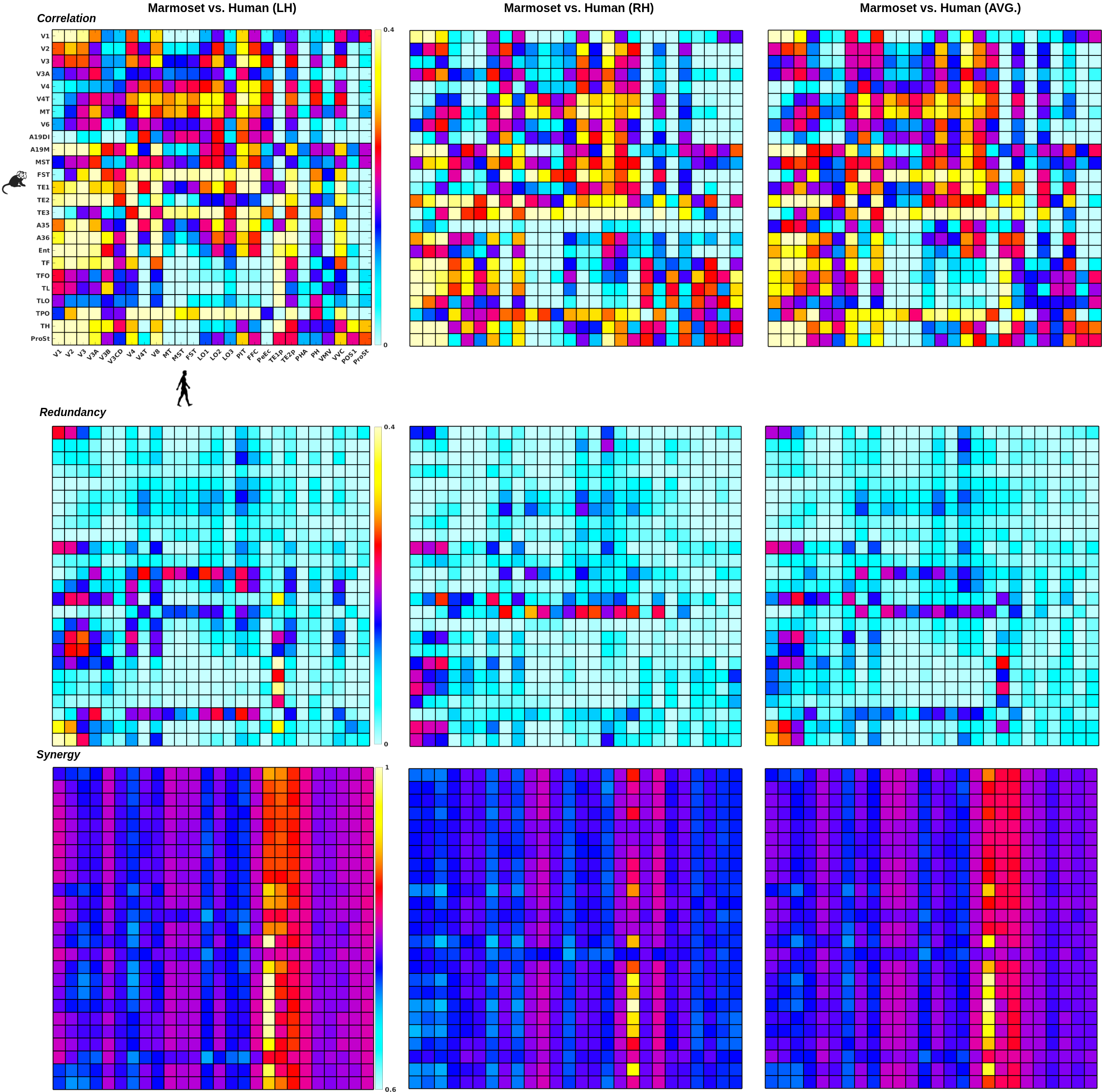}  
    \caption{Inter-species functional correspondences were measured using correlation, redundancy, and synergy in the aligned visual cortex of humans and marmosets. \textcolor{black}{The top row presents the correlation results for the left hemisphere (LH), right hemisphere (RH), and their average (AVG.), displayed from left to right. The second row shows redundancy results, highlighting the overlap and shared connections between the visual cortex regions across species. The final row illustrates the synergy measures, emphasizing the unique and complementary functional relationships between the human and marmoset visual cortices.}}
    \label{fig:2}
\end{figure}

\subsection{\textcolor{black}{Inter-species Functional Correspondences in Humans and Marmosets}}
In Fig.\ref{fig:2}, inter-species functional correspondences are presented. \textcolor{black}{First, we observed strong similarity patterns in correlation measures across the left and right hemispheres, as well as in their combined average. Second, specific connections identified through the synergy measure reveal distinct patterns not captured by correlation or redundancy. In particular, synergy across the primate visual cortex shows a markedly different distribution, highlighting unique inter-species relationships.} We find that humans and marmosets exhibit strong functional synergistic relationships in high-level visual regions. Notably, the PeEc region in humans shows the strongest functional synergy with marmoset occipitotemporal high-level visual regions, including areas A35, A36, Ent, TF, TFO, TL, TLO, and TH. \textcolor{black}{A high synergy value between humans and marmosets indicates that the two species share complementary and integrated patterns of information processing in the specific brain regions mentioned. This suggests that certain visual areas, particularly high-level visual regions, encode complex features in a coordinated and non-redundant manner across species. Such functional synergy supports the idea of conserved and evolutionarily shared mechanisms underlying advanced visual perception in primates, especially under naturalistic and socially relevant stimuli.}

\begin{figure}[!ht]
    \centering
    \includegraphics[width=0.96\textwidth, height=13cm]{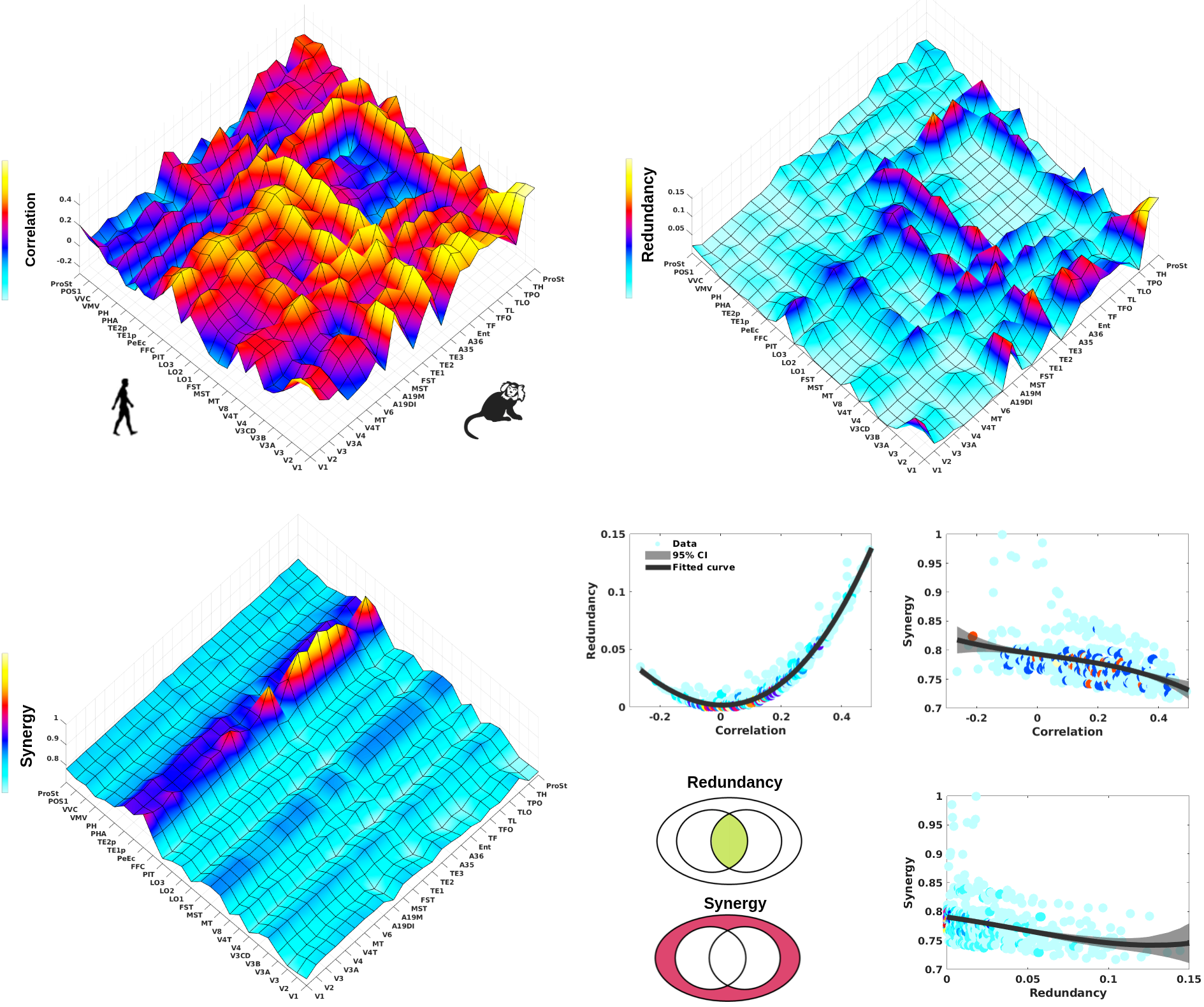} 
    \caption{\textcolor{black}{Functional correspondences using correlation, redundancy, and synergy were evaluated in the aligned visual cortex of humans and marmosets. The 3D surface shows average correlation, redundancy, and synergy measures across left and right hemisphere for both humans and marmosets, highlighting their similarities and differences. Additionally, the relationships among correlation, redundancy, and synergy were examined to demonstrate how these different metrics relate to one another across species.}}
    \label{fig:3}
\end{figure}

\textcolor{black}{To better represent the average functional correspondence measures of correlation, redundancy, and synergy, we present them as a 3D surface, enabling visual tracking of functional correspondence between humans and marmosets. First, we observe a high similarity between the correlation and redundancy measures, while a distinct pattern is evident in the synergy measures, as illustrated in Fig.\ref{fig:3}.}

Furthermore, to assess the relationships between the correlation, redundancy, and synergy measures, we examined the pairwise correlations among them. The similarity between correlation and redundancy is \( r = 0.81 \); between redundancy and synergy, it is \( r = -0.36 \); and between correlation and synergy, it is \( r = -0.42 \). These results indicate that correlation and redundancy share the strongest positive relationship compared to the other pairs.

\subsection{High-Order Hubs in the Visual Cortex of Humans and Marmosets}
\begin{figure}[!ht]
    \centering
    \includegraphics[width=\textwidth, height=13cm]{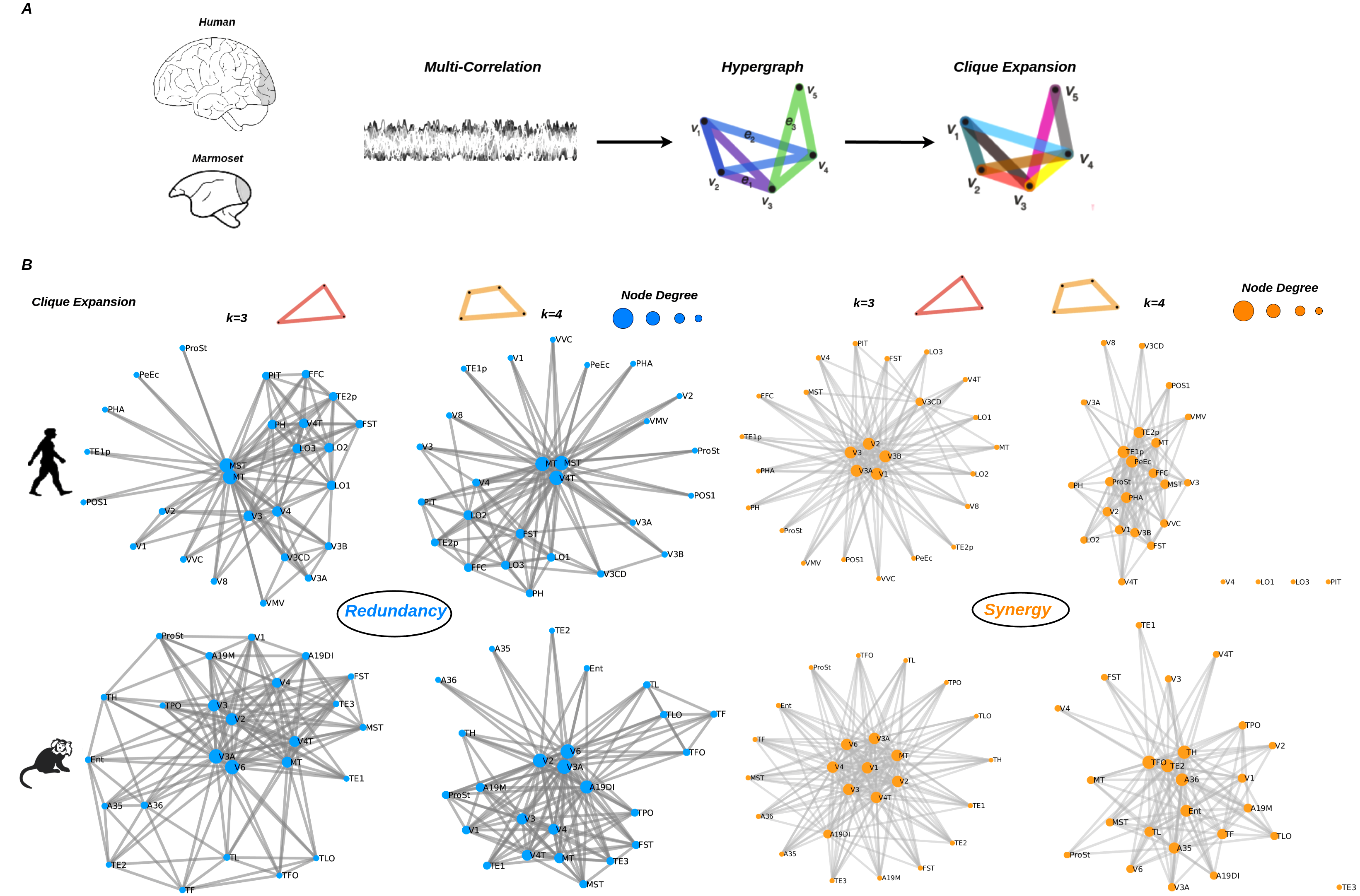} 
    \caption{High-order hubs in human and marmoset brains. \textbf{\textit{A.}} Multi-correlation was estimated from visual neural signals, followed by the construction of a hypergraph that captures beyond pairwise interactions. To facilitate the identification of high-order hubs in both species, the hypergraph was then transformed into a clique representation. \textbf{\textit{B.}} We examined third-order (k=3) and fourth-order (k=4) interactions in both species and visualized high-order functional connectivity using a force-layout graph. In this representation, redundancy and synergy connections are distinguished, while node size reflects node degree, highlighting key high-order hubs.}
    \label{fig:4}
\end{figure}

The high-order hubs identified in the visual cortex of humans and marmosets, as illustrated in Fig.\ref{fig:4}, were estimated using beyond pairwise interactions with interaction orders \( k=3 \) and \( k=4 \). To better interpret these interactions, we transformed hypergraph-based multi-region connections into pairwise representations. This approach enabled us to systematically identify critical high-order hubs in both species. Notably, redundancy measures preserved stable high-order hubs, while synergy measures revealed a dynamic transition from low- to high-level visual regions as more regions interacted, highlighting differences in information integration across the visual hierarchy.

From redundancy measures, we identified high-order hubs such as MST, MT, and V4T in the human visual cortex, while in marmosets, prominent hubs included V6, V3A, and V2. In synergy measures, for \( k=3 \), high-order hubs in humans included V1, V2, V3, V3A, and V3B, which then shifted to PeEc, TE1p, TE2p, and MT at \( k=4 \). Similarly, in marmosets, high-order hubs initially included V1, V2, V3, V3A, V4, V4T, V6, and MT, but transitioned to higher-level visual regions such as TFO, A36, TE2, TH, and Ent. \textcolor{black}{This transition aligns well with interspecies synergy measures, emphasizing the cross-species functional synergistic relationship in visual processing and reflecting the integrated mechanisms of visual information processing.}

\section{Discussion}
Functional correspondence across species offers valuable insights into the neural mechanisms of visual processing. Comparing how different species respond to similar stimuli reveals the evolutionary aspects of brain function, shared neural representations, and the roles of specific brain regions in visual perception. These insights bridge species-specific and universal visual processing pathways, thereby enhancing our understanding of brain functionality across primates~\cite{Hori21PNAS,Ghahremani16,Dureux23,Dureux2024}. In this study, we assessed functional correspondences in the visual cortex of humans and marmosets by employing both pairwise and beyond-pairwise metrics, including correlation, redundancy, and synergy. These measures quantified the information shared across species during exposure to identical movie-based visual stimuli. The findings provide deeper insights into information interactions within the primate visual cortex and offer further evidence of conserved visual processing mechanisms under naturalistic conditions.

\textcolor{black}{From the correlation and redundancy measures, we observed highly similar connection patterns within each species. Additionally, the functional synergy measures revealed particularly strong synergistic relationships among high-level visual regions, further emphasizing their integrative role in intra-species visual processing. For example, in the human visual cortex, the PeEc and TE1p exhibited especially strong functional synergy. Notably, both regions are selectively activated during facial recognition tasks, highlighting their involvement in processing face-related stimuli~\cite{Tsao2008,Rajimehr2009}. This suggests that these areas contribute not only to general visual perception but also play a specialized role in face processing, reinforcing the concept of dedicated face-selective networks in the human brain.}

\textcolor{black}{In the marmoset, similar functional synergy patterns were observed, particularly between TFO and A36. The TFO is known to integrate visual information from multiple cortical regions, acting as a hub that coordinates visual processing across different domains~\cite{Clery2020,Tsao2008,Rajimehr2009,Dureux2024}. This integration suggests that TFO plays a key role in linking low-level visual features with more complex visual representations (such as face recognition)~\cite{Dureux23jnc}}. A36, like PeEc in humans, is also a face-selective region in marmosets~\cite{Schaeffer20nc,Hung2015}. Therefore, the strong connection between TFO and A36 may underscore the integration of both low-level and high-level visual information required for complex visual tasks, such as object recognition and face perception.

Moreover, TFO's activation by condition-specific movies, such as the grasping hand condition~\cite{Zanini2023}, further emphasizes its role in processing dynamic, action-related stimuli. This suggests that TFO is not limited to static visual processing but also plays a critical role in perceiving and interpreting dynamic visual information, including movements and actions, which are essential for understanding social and environmental interactions~\cite{Dureux23jnc}.

By comparing the functional correspondences between these regions in humans and marmosets, we gain insight into the evolutionary conservation of visual processing mechanisms. The shared involvement of regions such as PeEc, TE1p, TFO, and A36 in both species highlights fundamental neural networks that support complex visual functions, particularly those related to face and action recognition. Understanding these functional similarities across species enhances our knowledge of how visual information is processed in the brain, revealing both shared and species-specific mechanisms that are critical for perception, cognition, and behavior.

From the interspecies functional correspondence analysis, we observed stronger functional synergy between the human PeEc and marmoset occipitotemporal visual regions. The PeEc in humans and the occipitotemporal regions in marmosets are both critical for visual perception, particularly for recognizing and processing faces~\cite{Tsao2008,Rajimehr2009,Dureux2024}. This alignment across species suggests that these brain areas play a conserved role in face perception, highlighting the evolutionary stability of face-selective processing mechanisms. Furthermore, the observed functional synergy between these regions underscores their joint involvement in integrating visual information from multiple sources-an essential process for performing complex visual tasks such as face recognition.

To further examine functional correspondence in the human and marmoset visual cortex, we found that redundancy measures preserved stable high-order hubs, suggesting a robust backbone of overlapping information processing that supports the hierarchical organization of the visual system across species. In contrast, synergy measures revealed a dynamic transition from low- to high-level visual regions as more areas interacted, underscoring an adaptive integration of specialized inputs that evolves with increasing network complexity. This duality highlights not only differences in how information is integrated across the visual hierarchy but also potential evolutionary divergences in cortical strategies for balancing stability and flexibility in perception.

\textcolor{black}{In summary, our findings underscore the marmoset’s value as a model for studying visual perception, due to its strong functional correspondence with the human visual cortex. Additionally, combining correlation, redundancy, and synergy measures offers a comprehensive and complementary perspective on the complex information processing that underlies visual function across species.}

\section{Data Availability}
The open-source fMRI data shared in previous research~\cite{Hori21PNAS} were used in this study. We thank Hori et al. for providing these data, which were crucial for our analysis.

\section{Author Contributions}
Qiang Li: Conceptualization; Formal analysis; Visualization; Writing - Original Draft. Ting Xu: Writing - Review \& Editing. Vince D. Calhoun: Writing - Review \& Editing.

\section{Funding Information}
This work received no financial support.

\printbibliography

@article{watanabe1960information,
  title={Information theoretical analysis of multivariate correlation},
  author={Watanabe, Satosi},
  journal={IBM Journal of research and development},
  volume={4},
  number={1},
  pages={66--82},
  year={1960},
  publisher={IBM}
}

@inproceedings{renyi1961measures,
  title={On measures of entropy and information},
  author={R{\'e}nyi, Alfr{\'e}d},
  booktitle={Proceedings of the Fourth Berkeley Symposium on Mathematical Statistics and Probability, Volume 1: Contributions to the Theory of Statistics},
  volume={4},
  pages={547--562},
  year={1961},
  organization={University of California Press}
}

@article{li2022functional,
  title={Functional connectivity inference from fMRI data using multivariate information measures},
  author={Li, Qiang},
  journal={Neural Networks},
  volume={146},
  pages={85--97},
  year={2022},
  publisher={Elsevier}
}

@article{li2022functionalnn,
author = {Li, Qiang and Ver Steeg, Greg and Malo, Jesús},
year = {2023},
month = {12},
pages = {127143},
title = {Functional connectivity via total correlation: Analytical results in visual areas},
volume = {571},
journal = {Neurocomputing}
}

@article{Varley2022MultivariateIT,
  title={Multivariate information theory uncovers synergistic subsystems of the human cerebral cortex},
  author={Thomas F. Varley and Maria Pope and Joshua Faskowitz and Olaf Sporns},
  journal={Communications Biology},
  year={2022},
  volume={6}
}

@article{Williams2010NonnegativeDO,
  title={Nonnegative Decomposition of Multivariate Information},
  author={Paul L. Williams and Randall D. Beer},
  journal={ArXiv},
  year={2010},
  eprint={1004.2515},
  archivePrefix={arXiv}
}

@misc{mediano2021extended,
      title={Towards an extended taxonomy of information dynamics via Integrated Information Decomposition}, 
      author={Pedro A. M. Mediano and Fernando E. Rosas and Andrea I Luppi and Robin L. Carhart-Harris and Daniel Bor and Anil K. Seth and Adam B. Barrett},
      year={2021},
      eprint={2109.13186},
      archivePrefix={arXiv}
}

@article{Barrett2011PracticalMO,
  title={Practical Measures of Integrated Information for Time-Series Data},
  author={Adam B. Barrett and Anil. K. Seth},
  journal={PLoS Computational Biology},
  year={2011},
  volume={7}
}

@article{luppi2022synergistic,
  title={A synergistic core for human brain evolution and cognition},
  author={Luppi, Andrea I and Mediano, Pedro AM and Rosas, Fernando E and Holland, Negin and Fryer, Tim D and O’Brien, John T and Rowe, James B and Menon, David K and Bor, Daniel and Stamatakis, Emmanuel A},
  journal={Nature Neuroscience},
  volume={25},
  number={6},
  pages={771--782},
  year={2022},
  publisher={Nature Publishing Group US New York}
}

@article{Lizier14fr,
author = {Lizier, Joseph},
year = {2014},
month = {08},
title = {JIDT: An Information-Theoretic Toolkit for Studying the Dynamics of Complex Systems},
volume = {1},
journal = {Frontiers in Robotics and AI}
}

@article{Hori21PNAS,
author = {Hori, Yuki and Cléry, Justine and Selvanayagam, Janahan and Schaeffer, David and Johnston, Kevin and Menon, Ravi and Everling, Stefan},
year = {2021},
month = {09},
title = {Interspecies activation correlations reveal functional correspondences between marmoset and human brain areas},
volume = {118},
journal = {Proceedings of the National Academy of Sciences of the United States of America}
}

@article{Qlinl24,
author = {Li, Qiang and Calhoun, Vince and Iraji, Armin},
year = {2024},
month = {01},
pages = {137624},
title = {Revealing complex functional topology brain network correspondences between humans and marmosets},
volume = {822},
journal = {Neuroscience Letters}
}

@article{Kell22iscience,
author = {Kell, Alexander and Bokor, Sophie and Jeon, You-Nah and Toosi, Tahereh and Issa, Elias},
year = {2022},
month = {12},
pages = {105788},
title = {Marmoset core visual object recognition behavior is comparable to that of macaques and humans},
volume = {26},
journal = {iScience}
}

@article{Mitchell14jn,
author = {Mitchell, Jude and Reynolds, John and Miller, Cory},
year = {2014},
month = {01},
pages = {1183-94},
title = {Active Vision in Marmosets: A Model System for Visual Neuroscience},
volume = {34},
journal = {The Journal of neuroscience : the official journal of the Society for Neuroscience}
}

@article{Schaeffer20nc,
author = {Schaeffer, David and Selvanayagam, Janahan and Johnston, Kevin and Menon, Ravi and Freiwald, Winrich and Everling, Stefan},
year = {2020},
month = {09},
pages = {4856},
title = {Face selective patches in marmoset frontal cortex},
volume = {11},
journal = {Nature Communications}
}

@article{Gilbert21nc,
author = {Gilbert, Kyle and Cléry, Justine and Gati, Sabiha and Hori, Yuki and Johnston, Kevin and Mashkovtsev, Alexander and Selvanayagam, Janahan and Zeman, Peter and Menon, Ravi and Schaeffer, David and Everling, Stefan},
year = {2021},
month = {11},
pages = {6608},
title = {Simultaneous functional MRI of two awake marmosets},
volume = {12},
journal = {Nature Communications}
}

@article{Albers12csf,
author = {Albers, DJ and Hripcsak, George},
year = {2012},
month = {06},
pages = {853-860},
title = {Estimation of time-delayed mutual information and bias for irregularly and sparsely sampled time-series},
volume = {45},
journal = {Chaos, solitons, and fractals}
}

@article{Glasser2016,
  author = {Glasser, Matthew F. and Coalson, Timothy S. and Robinson, Emma C. and Hacker, Carl D. and Harwell, John and Yacoub, Essa and Ugurbil, Kamil and Andersson, Jesper and Beckmann, Christian F. and Jenkinson, Mark and Smith, Stephen M.},
  title = {A multi-modal parcellation of human cerebral cortex},
  journal = {Nature},
  volume = {536},
  number = {7615},
  pages = {171--178},
  year = {2016}}

@book{Paxinos2012,
  author = {Paxinos, George},
  title = {The Marmoset Brain in Stereotaxic Coordinates},
  year = {2012},
  publisher = {Elsevier Academic Press}
}

@article{Drezner1995MultirelationaCA,
  title={Multirelation-a correlation among more than two variables},
  author={Zvi Drezner},
  journal={Computational Statistics \& Data Analysis},
  year={1995},
  volume={19},
  pages={283-292}
}

@article{Pickard23PLOS,
    author = {Pickard, Joshua AND Chen, Can AND Salman, Rahmy AND Stansbury, Cooper AND Kim, Sion AND Surana, Amit AND Bloch, Anthony AND Rajapakse, Indika},
    journal = {PLOS Computational Biology},
    title = {HAT: Hypergraph analysis toolbox},
    year = {2023},
    month = {06},
    volume = {19},
    pages = {1-7},
    number = {6}
}

@book{rabiner1978digital,
  title={Digital Processing of Speech Signals},
  author={Rabiner, Lawrence R. and Schafer, Ronald W.},
  year={1978},
  publisher={Prentice Hall},
  series={Signal Processing Series},
  pages={147--148},
  address={Upper Saddle River, NJ},
  isbn={0132136031}
}

@article{Zanini2023,
  author = {Zanini, A. and Dureux, A. and Selvanayagam, J. and others},
  title = {Ultra-high field fMRI identifies an action-observation network in the common marmoset},
  journal = {Commun Biol},
  volume = {6},
  pages = {553},
  year = {2023}
}

@article{Tsao2008,
  author = {Tsao, D. Y. and Moeller, S. and Freiwald, W. A.},
  title = {Comparing face patch systems in macaques and humans},
  journal = {Proc. Natl. Acad. Sci. U.S.A.},
  volume = {105},
  pages = {19514--19519},
  year = {2008}
}

@article{Rajimehr2009,
  author = {Rajimehr, R. and Young, J. C. and Tootell, R. B. H.},
  title = {An anterior temporal face patch in human cortex, predicted by macaque maps},
  journal = {Proc. Natl. Acad. Sci. U.S.A.},
  volume = {106},
  pages = {1995--2000},
  year = {2009}
}

@article{Hung2015,
  author = {Hung, C. C. and Yen, C. C. and Ciuchta, J. L. and Papoti, D. and Bock, N. A. and Leopold, D. A. and Silva, A. C.},
  title = {Functional mapping of face-selective regions in the extrastriate visual cortex of the marmoset},
  journal = {Journal of Neuroscience},
  volume = {35},
  number = {3},
  pages = {1160--1172},
  year = {2015}
}

@article{Clery2020,
author = {Cléry, Justine and Schaeffer, David and Hori, Yuki and Gilbert, Kyle and Schaeffer, Lauren and Gati, Joseph and Menon, Ravi and Everling, Stefan},
year = {2020},
month = {04},
pages = {116815},
title = {Looming and receding visual networks in awake marmosets investigated with fMRI},
volume = {215},
journal = {NeuroImage}
}

@article{Dureux2024,
  author = {Dureux, Audrey and Zanini, Alessandro and Everling, Stefan},
  title = {Mapping of facial and vocal processing in common marmosets with ultra-high field fMRI},
  journal = {Communications Biology},
  volume = {7},
  year = {2024},
  month = {03}
}

@article{Ghahremani16,
author = {Ghahremani, Maryam and Hutchison, R.Matthew and Menon, Ravi and Everling, Stefan},
year = {2016},
month = {07},
pages = {},
title = {Frontoparietal Functional Connectivity in the Common Marmoset},
volume = {27},
journal = {Cerebral Cortex}
}

@article{Dureux23,
author = {Dureux, Audrey and Zanini, Alessandro and Selvanayagam, Janahan and Menon, Ravi and Everling, Stefan},
year = {2023},
month = {07},
pages = {},
title = {Gaze patterns and brain activations in humans and marmosets in the Frith-Happé theory-of-mind animation task},
volume = {12},
journal = {eLife},
doi = {10.7554/eLife.86327}
}

@article{Dureux23jnc,
author = {Dureux, Audrey and Zanini, Alessandro and Everling, Stefan},
year = {2023},
month = {03},
pages = {},
title = {Face-Selective Patches in Marmosets Are Involved in Dynamic and Static Facial Expression Processing},
volume = {43},
journal = {The Journal of neuroscience : the official journal of the Society for Neuroscience}
}
\end{document}